\documentclass[letterpaper,twocolumn,prl,aps,superscriptaddress,floatfix]{revtex4}

\usepackage[latin9]{inputenc}
\usepackage{amsmath}
\usepackage{amssymb}
\usepackage{graphicx}
\usepackage{esint}
\usepackage{times}
\usepackage{pdfpages}

\usepackage[unicode=true,
bookmarks=true,bookmarksnumbered=false,bookmarksopen=false,
breaklinks=false,pdfborder={0 0 1},backref=false,colorlinks=true]
{hyperref}
\hypersetup{
    linkcolor=magenta, urlcolor=blue, citecolor=blue, pdfstartview={FitH}, hyperfootnotes=false, unicode=true}

\makeatletter

\pdfpageheight\paperheight
\pdfpagewidth\paperwidth

\@ifundefined{textcolor}{}
{%
    \definecolor{BLACK}{gray}{0}
    \definecolor{WHITE}{gray}{1}
    \definecolor{RED}{rgb}{1,0,0}
    \definecolor{GREEN}{rgb}{0,1,0}
    \definecolor{BLUE}{rgb}{0,0,1}
    \definecolor{CYAN}{cmyk}{1,0,0,0}
    \definecolor{MAGENTA}{cmyk}{0,1,0,0}
    \definecolor{YELLOW}{cmyk}{0,0,1,0}
}

\usepackage{xcolor}
\usepackage{soul}

\newcommand{\bra}[1]{\ensuremath{\left\langle#1\right|}}
\newcommand{\ket}[1]{\ensuremath{\left|#1\right\rangle}}
\definecolor{blue}{rgb}{0,0,1}
\definecolor{red}{rgb}{1,0,0}
\definecolor{green}{rgb}{0,1,0}

\makeatother

\begin{document}

\title{High-fidelity, high-scalability two-qubit gate scheme for superconducting qubits}

\author{Yuan Xu}
\email{xuy5@sustech.edu.cn}
\affiliation{Shenzhen Institute for Quantum Science and Engineering, Southern University of Science and Technology, Shenzhen, Guangdong, China}
\affiliation{Guangdong Provincial Key Laboratory of Quantum Science and Engineering, Southern University of Science and Technology, Shenzhen, Guangdong, China}
\affiliation{Shenzhen Key Laboratory of Quantum Science and Engineering, Southern University of Science and Technology, Shenzhen, Guangdong, China}

\author{Ji Chu}
\affiliation{National Laboratory of Solid State Microstructures, Department of Physics, Nanjing University, Nanjing, Jiangsu, China}

\author{Jiahao Yuan}
\affiliation{Institute for Quantum Science and Engineering and Department of Physics, Southern University of Science and Technology, Shenzhen, Guangdong, China}

\author{Jiawei Qiu}
\affiliation{Institute for Quantum Science and Engineering and Department of Physics, Southern University of Science and Technology, Shenzhen, Guangdong, China}

\author{Yuxuan Zhou}
\affiliation{Institute for Quantum Science and Engineering and Department of Physics, Southern University of Science and Technology, Shenzhen, Guangdong, China}

\author{Libo Zhang}
\affiliation{Shenzhen Institute for Quantum Science and Engineering, Southern University of Science and Technology, Shenzhen, Guangdong, China}
\affiliation{Guangdong Provincial Key Laboratory of Quantum Science and Engineering, Southern University of Science and Technology, Shenzhen, Guangdong, China}
\affiliation{Shenzhen Key Laboratory of Quantum Science and Engineering, Southern University of Science and Technology, Shenzhen, Guangdong, China}

\author{Xinsheng Tan}
\email{tanxs@nju.edu.cn}
\affiliation{National Laboratory of Solid State Microstructures, Department of Physics, Nanjing University, Nanjing, Jiangsu, China}

\author{Yang Yu}
\affiliation{National Laboratory of Solid State Microstructures, Department of Physics, Nanjing University, Nanjing, Jiangsu, China}

\author{Song Liu}
\affiliation{Shenzhen Institute for Quantum Science and Engineering, Southern University of Science and Technology, Shenzhen, Guangdong, China}
\affiliation{Guangdong Provincial Key Laboratory of Quantum Science and Engineering, Southern University of Science and Technology, Shenzhen, Guangdong, China}
\affiliation{Shenzhen Key Laboratory of Quantum Science and Engineering, Southern University of Science and Technology, Shenzhen, Guangdong, China}

\author{Jian Li}
\email{lij33@sustech.edu.cn}
\affiliation{Shenzhen Institute for Quantum Science and Engineering, Southern University of Science and Technology, Shenzhen, Guangdong, China}
\affiliation{Guangdong Provincial Key Laboratory of Quantum Science and Engineering, Southern University of Science and Technology, Shenzhen, Guangdong, China}
\affiliation{Shenzhen Key Laboratory of Quantum Science and Engineering, Southern University of Science and Technology, Shenzhen, Guangdong, China}

\author{Fei Yan}
\email{yanf2020@mail.sustech.edu.cn}
\affiliation{Shenzhen Institute for Quantum Science and Engineering, Southern University of Science and Technology, Shenzhen, Guangdong, China}
\affiliation{Guangdong Provincial Key Laboratory of Quantum Science and Engineering, Southern University of Science and Technology, Shenzhen, Guangdong, China}
\affiliation{Shenzhen Key Laboratory of Quantum Science and Engineering, Southern University of Science and Technology, Shenzhen, Guangdong, China}

\author{Dapeng Yu}
\affiliation{Shenzhen Institute for Quantum Science and Engineering, Southern University of Science and Technology, Shenzhen, Guangdong, China}
\affiliation{Guangdong Provincial Key Laboratory of Quantum Science and Engineering, Southern University of Science and Technology, Shenzhen, Guangdong, China}
\affiliation{Shenzhen Key Laboratory of Quantum Science and Engineering, Southern University of Science and Technology, Shenzhen, Guangdong, China}

\begin{abstract}
High-quality two-qubit gate operations are crucial for scalable quantum information processing. Often, the gate fidelity is compromised when the system becomes more integrated. Therefore, a low-error-rate, easy-to-scale two-qubit gate scheme is highly desirable. Here, we experimentally demonstrate a new two-qubit gate scheme that exploits fixed-frequency qubits and a tunable coupler in a superconducting quantum circuit. The scheme requires less control lines, reduces crosstalk effect, simplifies calibration procedures, yet produces a controlled-$Z$ gate in 30~ns with a high fidelity of 99.5\%, derived from the interleaved randomized benchmarking method. Error analysis shows that gate errors are mostly coherence limited. Our demonstration paves the way for large-scale implementation of high-fidelity quantum operations.
\end{abstract}
\maketitle
\vskip 0.5cm


Quantum information processor architectures are scaling up at a fast pace, entering the Noisy Intermediate-Scale Quantum (NISQ) era~\cite{Preskill2018, Arute2019, Otterbach2017, Omran2019, Song2019, Yan2019}. The prospect of demonstrating quantum advantages with NISQ devices relies critically on continuing extending the system size without compromising the quality of quantum operations. Currently, two-qubit gate operation is the performance bottleneck in various modalities~\cite{Zajac2018, Wright2019, He2019, Kjaergaard2020}, and it generally deteriorates as more qubits, and hence more control lines, are integrated together. Obviously, more control lines introduce additional decohering channels, exacerbates crosstalk, and adds to the complexity of calibration procedures. Therefore, a high-fidelity yet easy-to-scale two-qubit gate scheme is the key to scalable quantum information processing.

For high-scalability two-qubit gates, two ingredients are highly desirable.
First, the use of a tunable coupler between qubits has been proven effective in resolving the problem of frequency crowding, suppressing residual coupling, and enabling fast and high-fidelity two-qubit gates~\cite{Chen2014,Yan2018,Arute2019,Mundada2019,Li2019,Han2020,Foxen2020}. However, humongous calibration efforts are required for precise control, especially when both qubits and couplers are tunable and sensitive to crosstalk~\cite{Arute2019}. The iterative and exquisite system tuning up adds instability to processor performance, hindering further scaling up.
Second, fixed-frequency qubits can drastically simplify the system, require less control lines and have better coherence in general. Previous experiments have demonstrated these advantages with nontunable superconducting qubits made with single Josephson junction~\cite{Chow2011, McKay2016, Caldwell2018, Bengtsson2019, Blok2020}. However, two-qubit interactions in these schemes are activated by parametrically driving the system, an inherently slow process that is prone to decohering errors. An ideal solution is a two-qubit gate scheme that takes advantage of both fixed-frequency qubits and tunable coupler, while maintaining high fidelity.

In this Letter, we experimentally demonstrate a new two-qubit gate scheme, compatible with fixed-frequency qubits, in a superconducting quantum circuit. Effective longitudinal qubit-qubit coupling can be adjusted by a single control parameter of a tunable coupler. With proper choice of the idling point, the system has a residual coupling strength as small as 20\,kHz. Taking advantage of enhanced adiabaticity due to strong qubit-coupler coupling ($>\!100$\,MHz), we realize a fast (30\,ns) and high-fidelity (99.5\%) adiabatic controlled-$Z$ (CZ) gate. Error analysis from separate measurements shows that the fidelity is limited mostly by decoherence. Moreover, our scheme is intrinsically robust against crosstalk and requires only a simple calibration sequence, promising better scalability in practice.

\begin{figure}
    \includegraphics{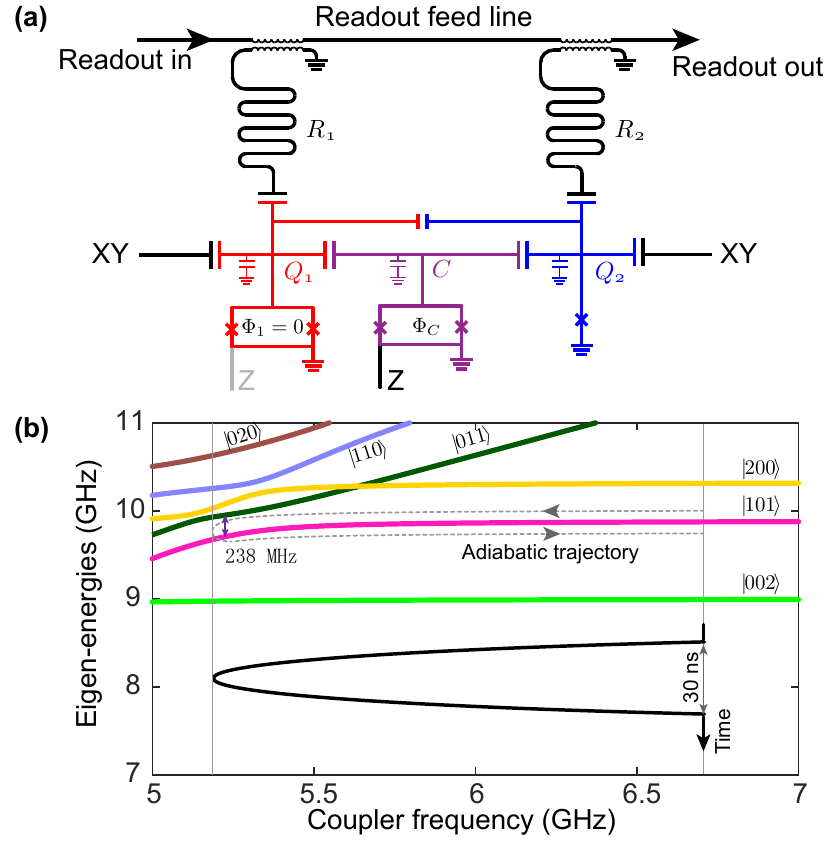}
    \caption{Device schematic and concept of the adiabatic CZ gate. \textbf{(a)} Simplified circuit schematic of the experimental sample. The two qubits have a Xmon design, and the central tunable coupler is also a transmon-type qubit. The all-capacitive-coupling architecture allows convenient engineering of the nearest-neighbor and the weaker next-nearest-neighbor couplings. Single-qubit operations are implemented with local $XY$ control lines. Two-qubit gates are implemented by modulating the magnetic flux threading the coupler's superconducting quantum interference device loop $\Phi_C$ with the local $Z$ control line. The first qubit is treated as an equivalent of a fixed-frequency qubit by setting its loop flux $\Phi_1$ to zero throughout the experiment. Two $\lambda/4$ resonators, coupled to the same transmission line, are used for reading out the qubit states simultaneously. \textbf{(b)} System eigen-energies as a function of the coupler frequency. Only states in the two-excitation manifold are shown. The adiabatic CZ gate is realized by a 30-ns flux pulse applied to the coupler. The pulse assumes a simple half-period cosine shape. The gray dashed line indicates the adiabatic trajectory of an initial $\ket{101}$ state, which follows the pink level. The smallest gap between the pink state and the other states is about 238~MHz.}
    \label{fig1}
\end{figure}

Our experiment is performed on a superconducting quantum circuit which consists of two Xmon qubits ($Q_1$, $Q_2$)~\cite{Barends2013Xmon} and a transmon-type~\cite{koch2007transmon} tunable coupler ($C$) in between, as shown in Fig.~\ref{fig1}(a). Note that the first qubit is made tunable for another experimental purpose~\cite{Qiu2020}. Throughout this Letter, it is biased at its maximum frequency, and can be treated as an equivalent of a fixed-frequency qubit. The system Hamiltonian can be expressed as
\begin{eqnarray}
\label{H1}
H/\hbar &=& \sum_{i=1,2,c}{\omega_i \, a_i^+ a_i + \frac{\alpha_i}{2} \, a_i^+ a_i^+ a_i a_i} \notag\\
&+&  \sum_{i\ne j}{g_{ij} \left( a_i^+ a_j + a_i a_j^+ \right)},
\end{eqnarray}
where $a_i^+$ and $a_i$ are corresponding creation and annihilation operators. $\omega_1/2\pi = 5.27$\,GHz and $\omega_2/2\pi = 4.62$\,GHz are the qubit frequencies. The coupler frequency $\omega_c$ is flux dependent, and is biased at $\omega_c/2\pi=6.74$\,GHz during idling and single-qubit gate periods. The corresponding anharmonicities are $\alpha_1/2\pi = -210$\,MHz, $\alpha_2/2\pi = -240$\,MHz, and $\alpha_c/2\pi = -370$\,MHz. To speed up the two-qubit gate while minimizing unwanted transitions, our design features enhanced coupling parameters. That is, $g_{12}/2\pi = 12$\,MHz (between qubits), $g_{1c}/2\pi = 122$\,MHz and $g_{2c}/2\pi = 105$\,MHz (between qubit and coupler), much stronger than the conventional Xmon design~\cite{Barends2013Xmon}. More details about the device and experimental setup can be found in the Supplemental Material~\cite{Supplement}.

To illustrate how the adiabatic CZ gate is implemented, we may rewrite the system Hamiltonian in Eq.~(\ref{H1}) using a generic form in its energy eigenbases ($\ket{Q_1,C,Q_2}$, labeled by the approximate bare states when the coupler is far detuned):
\begin{eqnarray}
\label{H2}
H\rq{} / \hbar &=& \tilde{\omega}_1 \ket{100}\bra{100} + \tilde{\omega}_2 \ket{001}\bra{001} \notag\\
&+& (\tilde{\omega}_1 + \tilde{\omega}_2 + \chi_{12}) \ket{101}\bra{101},
\end{eqnarray}
after truncation to the computational subspace. The eigenenergies $\tilde{\omega}_1$, $\tilde{\omega}_2$, and $\chi_{12}$ are all $\omega_c$ dependent. $\chi_{12}$ denotes the effective longitudinal coupling between qubits and is responsible for generating the entanglement. Finite $\chi_{12}$ is a consequence of interactions among higher levels, which can be relatively strong in transmon-type qubits due to their weak anharmonicity.
The energy levels adjacent to $\ket{101}$ are plotted in Fig.~\ref{fig1}(b) as a function of $\omega_c$. In our two-qubit gate scheme, we adiabatically adjust the coupler from an idling bias to a region where the bare state $\ket{101}$ interacts more strongly with other levels and then back to the original bias. Nonzero $\chi_{12}$ during this process leads to a controlled-phase operation or a CZ gate if the total accumulated phase is $\pi$.

Note that when the coupler frequency is tuned down by the CZ pulse, the bare frequency of $\ket{011}$ has already crossed the level of $\ket{101}$ (i.e. $\omega_c<\omega_1$). However, we emphasize that the strong coupling between $\ket{101}$ and $\ket{011}$ is not the sole cause of $\chi_{12}$, because $\ket{011}$ alone would have exactly the same level-pushing effect to $\ket{101}$ as $\ket{010}$ would do to $\ket{100}$, leading to a trivial single-qubit frequency shift.
In fact, the nontrivial interaction mainly arises from couplings between $\ket{101}$ and second-excited states, which are $\ket{200}$, $\ket{020}$, and $\ket{002}$. These interactions may become much stronger when their interaction-mediating states, $\ket{110}$ and $\ket{011}$, are tuned close to them. In the specific device, $\ket{020}$ leads the contribution to $\chi_{12}$, because $\ket{020}$ gets much closer to $\ket{101}$ in the interested regime than other second-excited states, given relatively larger anharmonicity of the coupler. The scheme may further benefit from optimizing the device parameters~\cite{Chu2020}.

The adiabatic process is supposed to be slow enough to avoid unwanted transitions, e.g., leakage to noncomputational states~\cite{DiCarlo2009}. In the conventional fixed-coupling architecture, the limit on the adiabatic CZ gate speed is set by the qubit-qubit coupling strength. In our device, with the introduction of a tunable coupler and its strong couplings to qubits, nonadiabatic effect can be effectively suppressed. The minimum gap between $\ket{101}$ and other states, a key factor in determining adiabaticity, is about 238\,MHz, much greater than that in the conventional scheme. Also, we find that our scheme adds robustness in adiabaticity against parameter instability and pulse distortion~\cite{Supplement}. Other non-adiabatic approaches that take advantage of the interference effect can facilitate the gate speed~\cite{Rol2019, Li2019Realisation, Barends2019Diabatic}, but may become sensitive to pulse distortion, adding instability to gate performance.


\begin{figure}
    \includegraphics{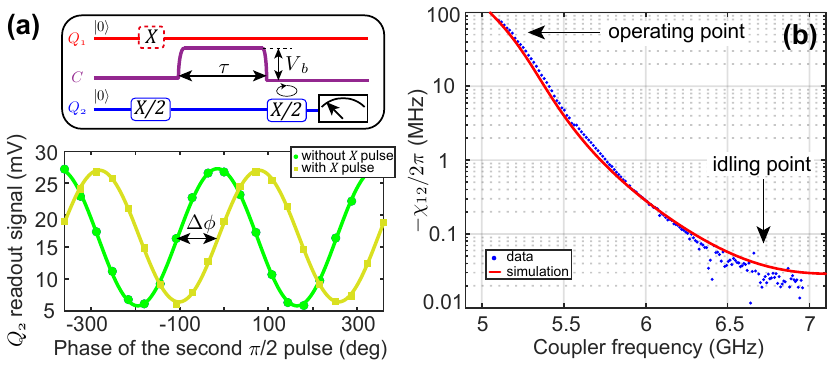}
    \caption{Tunability of the longitudinal coupling. \textbf{(a)} In the top panel is the control sequence that performs a Ramsey-like sequence on $Q_2$ conditioning on the state of $Q_1$ in order to extract the longitudinal coupling. Between the two $\pi/2$ pulses applied to $Q_2$, a squarelike flux pulse with amplitude $V_b$ and duration $\tau$ is applied to the coupler. The phase shift induced by this pulse is captured by $Q_2$ with the Ramsey-like sequence. In our experiment, the phase of the last $\pi/2$ pulse is a varying parameter (indicated by the circling arrow) so that a full oscillation can be resolved, as shown in the bottom panel. Data (markers) are fitted (solid lines) by a sinusoidal function to extract the differential phase ($\Delta\phi$) between the cases of $Q_1$ being at the ground (green) or excited (yellow) state.
    \textbf{(b)} Effective longitudinal coupling strength $\chi_{12}$ as a function of the coupler frequency. $\chi_{12}$ can be derived from the previous results in (a) by the relation $\Delta\phi=\chi_{12} \tau$. The experiment is repeated with different pulse amplitudes $V_b$ and fixed $\tau=50$ns ($500$ns) for larger (smaller) $V_b$. 
    Together with separately measured coupler spectrum~\cite{Supplement}, we obtain $\chi_{12}(\omega_c)$ (blue dots), which is in good agreement with numerical results (red line). We choose $\omega_c/2\pi = 6.74$~GHz as the idling point in subsequent two-qubit gate experiments, due to the small residual coupling (20~kHz). Note that the negative of $\chi_{12}$ is plotted here.
    }
    \label{fig2}
\end{figure}

In our experiment, we first measure the longitudinal coupling strength at different coupler frequencies, i.e., $\chi_{12}(\omega_c)$, from a conditional Ramsey-like experiment, as detailed in Fig.~\ref{fig2}. The dynamic range of the longitudinal coupling strength spans more than 3 orders of magnitude, from 20~kHz to 100~MHz, enabling fast two-qubit gate operations as well as small residual coupling.
The results are in good agreement with numerical simulation using our device parameters. Notably, since the two qubits have relatively large detuning ($|\omega_1-\omega_2|>|\alpha_1|$), there is no working bias such that the coupling can be turned off completely. This is different from the case when the two qubits are prepared to be near resonance~\cite{Yan2018,Arute2019}. However, we can still find a minimum coupling that is small enough (20~kHz) for practical applications.

\begin{figure}
    \includegraphics{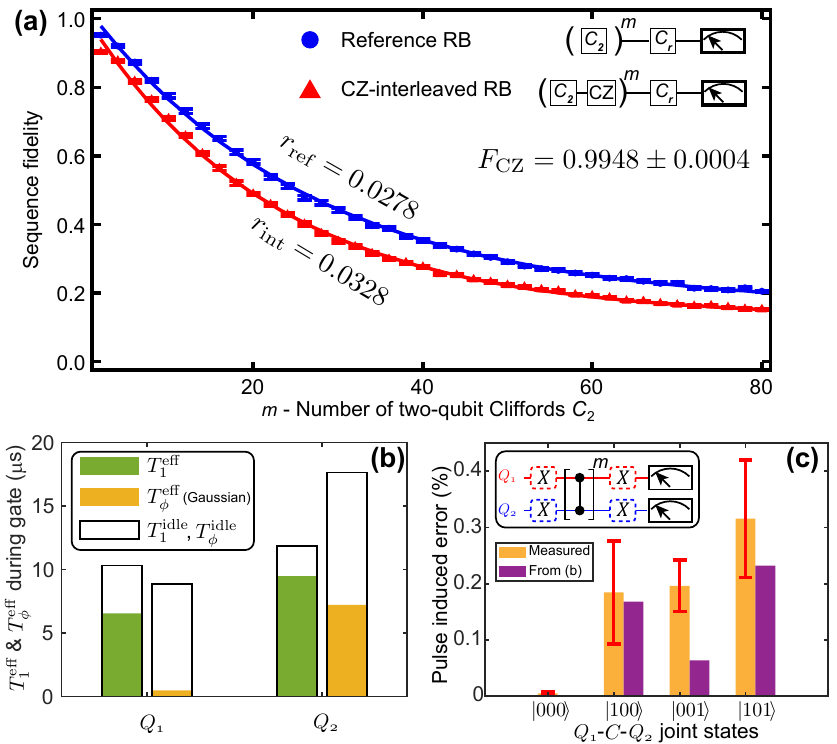}
    \caption{Fidelity analysis of the adiabatic CZ gate. 
    \textbf{(a)} Normalized sequence fidelity measured as a function of the number of Cliffords for both the reference (blue) and interleaved (red) RB experiments, with the sequences shown in the inset. Error bars on the data points are the standard deviations from the mean. We first obtain the decay constants, $p_\mathrm{ref}$ and $p_\mathrm{int}$, from an exponential fit, $F = Ap^m +B$ (solid lines), and then derive the error rate per Clifford, $r_\mathrm{ref}$ and $r_\mathrm{int}$ from $r = \frac{3}{4}\left(1-p\right)$. We also extract the CZ gate error from $r_\mathrm{CZ} = \frac{3}{4}\left(1-p_\mathrm{int}/p_\mathrm{ref}\right)$ and finally the CZ gate fidelity $F_\mathrm{CZ} = 1-r_\mathrm{CZ}$.
    The uncertainty of the fidelity is determined by bootstrapping. 
    \textbf{(b)} Effective energy relaxation time $T_1^\mathrm{eff}$ (green bar) and pure dephasing (Gaussian decay) time $T_\phi^\mathrm{eff}$ (orange bar) during the adiabatic CZ gate. The results are calculated by averaging over $\omega_c$, weighted by the actual pulse shape (seeSupplemental Material~\cite{Supplement} for details). The blank outlines indicate measured characteristic times when the coupler is at the idling point. 
    \textbf{(c)} Pulse-induced transitional errors by the CZ gate for different joint states (orange bars). Inset: control sequence. The qubits are prepared at the four different joint states with conditional $\pi$ pulses. In each case, a varying number ($m$) of identical CZ gates are repeatedly applied. The final decay curves (versus $m$) are compared with a reference case in which CZ gates are replaced by identity gates of the same length, so that additional transition rates or errors due to the pulsing can be extracted (see~\cite{Supplement} for details). The additional errors from the shortening of energy relaxation times during pulse are calculated by comparing the effective and idling $T_1$ times in (b) and shown for comparison (purple bars). }
    \label{fig3}
\end{figure}

Next, we calibrate the adiabatic CZ gate.
As shown in Fig.~\ref{fig1}(b), the 30-ns flux pulse assumes a half-period cosine shape, with rising and falling edges smooth enough for adiabatic evolution at this timescale. A conditional Ramsey experiment similar to the one shown in Fig.~\ref{fig2}(a) is used for calibrating the amplitude of the flux pulse, the only free parameter at this step. We obtain a CZ gate when the conditional phase shift satisfies $\Delta\phi = \pi$ and also find out the parasitic single-qubit phases, later to be compensated by virtual-$Z$ gates~\cite{McKay2017}. For subsequent randomized benchmarking (RB) experiments, these parameters are further optimized using the RB results as the cost function~\cite{Kelly2014}.

We assess our gate performance by the conventional Clifford-based RB method~\cite{RBSQProtocol, RBMultiQProtocol, RBInterleaved}, which measures the decay of the ground state probability (sequence fidelity) as a function of the number of two-qubit Cliffords $m$ for both the reference and CZ-interleaved RB experiments, as shown in Fig.~\ref{fig3}(a). With exponential fit, we obtain the average error per Clifford $r_\mathrm{ref} = 0.0278$ and $r_\mathrm{int} = 0.0328$.
By comparing the exponential decay constants of the two traces, we extract the CZ gate fidelity $F_\mathrm{CZ} = 1 -  \frac{3}{4}\left(1-p_\mathrm{int}/p_\mathrm{ref}\right) = 0.9948\pm0.0004$. We note that the difference in the trace offsets may result from leakage~\cite{Supplement}. Since the ratio $r_\mathrm{CZ}/r_\mathrm{ref}\approx0.18$ is small, there are possibilities of significant systematic variations in the interleaved RB results~\cite{Epstein2014}. As a consistency check, we rederive the CZ gate fidelity by subtracting single-qubit errors from the reference RB result using the relation $r_\mathrm{ref} = 1.5 \,r_\mathrm{CZ} + 8.25 \,r_\mathrm{1q}$, where $r_\mathrm{1q}=0.0013$~\cite{Supplement}. The resulting error rate, 1.14\%, sets an upper bound of the CZ gate fidelity.

To estimate the decoherence error, we first obtain the effective energy relaxation time $T_1^\mathrm{eff}$ and pure dephasing time $T_\phi^\mathrm{eff}$ during the CZ gate [Fig.~\ref{fig3}(b)].
Obviously, $T_1^\mathrm{eff}$ and $T_\phi^\mathrm{eff}$ are lower than their counterparts during idling periods ($T_1^\mathrm{idle}$ and $T_\phi^\mathrm{idle}$). In particular, the higher-frequency $Q_1$ dephases drastically faster ($T_{\phi,Q_1}^\mathrm{eff}\approx0.5~\mu$s), a consequence of the stronger interaction between the qubit and the less coherent coupler during the CZ pulse. The (Gaussian) pure dephasing from $Q_1$ contributes $\frac{1}{3} (\tau_\mathrm{gate} / T_{\phi,Q_1}^\mathrm{eff})^2  = 0.12\%$ to gate errors~\cite{OMalley2015}, where $\tau_\mathrm{gate}=30$~ns is the pulse duration. The dephasing contribution from $Q_2$ ($\ll0.01\%$) is negligible.
The $T_1$ contribution to gate errors can be estimated by an empirical relation, $\frac{1}{3} \left( \frac{\tau_\mathrm{gate}}{\bar{T}_{1,Q_1}^\mathrm{eff}} + \frac{\tau_\mathrm{gate}}{\bar{T}_{1,Q_2}^\mathrm{eff}}  +  \frac{\tau_\mathrm{spacing}}{\bar{T}_{1,Q_1}^\mathrm{idle}} + \frac{\tau_\mathrm{spacing}}{\bar{T}_{1,Q_2}^\mathrm{idle}} \right) = 0.28\%$, where $\tau_\mathrm{spacing}=4$~ns is the interpulse spacing.
Therefore, decoherence, including both $T_1$ and $T_\phi$ processes, accounts for about 77\% of the total gate error, while pulse-induced nonadiabatic effects account for 23\%, consistent with the numerical simulation result~\cite{Supplement}.

To validate our analysis, we perform a separate experiment, measuring the pulse-induced transitional errors on each of the four joint eigenstates, as detailed in Fig.~\ref{fig3}(c) and the Supplemental Material~\cite{Supplement}.
Note that the extracted transition rates per gate denote the additional errors caused by the CZ pulse, compared to the identity operation. These errors include two parts, additional energy relaxation during gate and transitions caused by the pulse-induced dynamic effect, in this case, the nonadiabatic effect.
The average difference between these rates and the $T_1$ contribution (orange versus purple bars in Fig.~\ref{fig3}(c)), $(0.06\pm0.06)\%$, is hence the non-adiabatic errors, consistent with the above analysis.

\begin{figure}
    \includegraphics{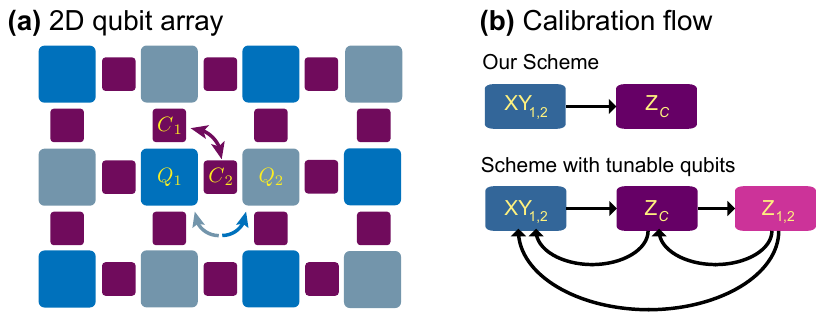}
    \caption{Considerations on crosstalk and calibration. 
    \textbf{(a)} Layout of a surface-code-compatible qubit array implementing our two-qubit gate scheme. Blue and gray squares denote qubits with different frequencies. The single-qubit ($XY$) control signals of two neighboring qubits ($Q_1$ and $Q_2$) have different frequency components, so their crosstalk has little influence on qubits. Although the two-qubit ($Z$) control signals applied to the tunable couplers (purple squares) share the same bandwidth, the flux crosstalk between neighboring couplers ($C_1$ and $C_2$) does not necessarily lead to adverse effect, because when the maximum frequency of the coupler is designed at the idling point (minimal residual coupling), an idling coupler becomes insensitive to flux, and also the longitudinal coupling is insensitive to the coupler frequency. Note that neighboring couplers can only be operated in turn in regular quantum circuits. 
    \textbf{(b)} Comparison of the calibration flow. In our adiabatic CZ gate scheme, the qubit-$XY$ signals (for single-qubit gate) are first calibrated before calibrating the coupler-$Z$ signal (for two-qubit gate). In comparison, the iSWAP-like gate scheme~\cite{Arute2019} which requires stringent resonance condition between qubits demands iterative calibration procedures. For example, the qubit and coupler idling biases and control pulses have to be recalibrated once the other one is changed. The process may be further complicated by crosstalk and signal distortion.}
    \label{fig4}
\end{figure}

Finally, we discuss the scalability of our scheme from the perspective of crosstalk and calibration. Consider a 2D qubit array for implementing surface code~\cite{Fowler2014}, as shown in Fig.~\ref{fig4}(a). With the problem of frequency crowding addressed by tunable couplers, we may pattern the qubit array with an interleaved frequency setup. Such an arrangement provides robustness against the $XY$-line crosstalk between neighboring qubits, because of the ineffectiveness of driving a qubit with a frequency-detuned signal.
More importantly, our scheme is also intrinsically robust against the notorious $Z$-line crosstalk. Given that the maximum frequency of the coupler can be designed at the idling point (minimum residual coupling), the longitudinal coupling becomes doubly insensitive to flux variations, since both $\chi_{12}(\omega_c)$ and $\omega_c(\Phi_C)$ are at first order insensitive points. From our device parameters, it is estimated that a $10\%$ flux crosstalk from the neighboring $Z$-line only incurs an additional coupling less than 1~kHz.
The robustness against nearest-neighbor crosstalk in both types of control lines can provide more flexibility in signal routing in large-scale devices.

Compared to the state-of-the-art result using tunable qubits with tunable coupler~\cite{Arute2019}, the calibration procedures for finding the optimized system and control parameters in our scheme are drastically simplified, as shown in Fig.~\ref{fig4}(b). Since the qubits are fixed frequency and the couplers are insusceptible to crosstalk, the calibration process does not require iterative (cross) tuning-ups or complicated check procedures. Single- and two-qubit control parameters are calibrated separately in turn. The simplicity of our scheme not only reduces calibration procedures, but also adds stability to performance by lowering the probability of failure or bad events, enabling reliable chip-scale automated calibration.

To conclude, we experimentally demonstrate a new type of adiabatic CZ gate with fixed-frequency qubits and a tunable coupler in a superconducting quantum circuit. With a large on-off ratio ($>\!1000$) of the effective coupling adjustable by the coupler frequency (flux), we achieve small residual coupling (20~kHz) and fast CZ gate (30~ns). A high gate fidelity of 99.5\% is obtained from interleaved randomized benchmarking, with error analysis showing mostly coherence-limited gate error. The gate performance may further benefit from optimized pulse shape for faster adiabatic process~\cite{FastCZMartinis} and from coherence improvement with new material platform~\cite{Place2020}.
Also, our scheme is easy to scale due to its intrinsic robustness against crosstalk and a simple calibration flow. This high-fidelity, high-scalability two-qubit gate scheme promises reproducibly high-quality quantum operations in future large-scale quantum information processors.

\begin{acknowledgments}
This work was supported by the Key-Area Research and Development Program of Guangdong Province (Grant No. 2018B030326001), the National Natural Science Foundation of China (Grant No. U1801661), the Guangdong Provincial Key Laboratory (Grant No. 2019B121203002), the Guangdong Innovative and Entrepreneurial Research Team Program (Grant No. 2016ZT06D348), the Natural Science Foundation of Guangdong Province (Grant No. 2017B030308003) and the Science, Technology and Innovation Commission of Shenzhen Municipality (Grants No. JCYJ20170412152620376, No. KYTDPT20181011104202253). J.L. acknowledges support from the National Natural Science Foundation of China (Grant No. 11874065) and the Natural Science Foundation of Hunan Province (Grant No. 2018JJ1031). Y.X. acknowledges support from the Youth Project of National Natural Science Foundation of China (Grant No. 11904158). Y.Y. acknowledges support from the National Key Research and Development Program of China (Grant No. 2016YFA0301802).
\end{acknowledgments}

Y. X. and J. C. contributed equally to this work.

\textit{Note added.}-- We notice concurrent development of a similar work to implement conditional-phase gates with a tunable coupler~\cite{Collodo2020}.

%


\includepdf[pages={1,{},2,{},3,{},4,{},5,{},6,{},7,{},8,{},9}]{Supplement.pdf}

\end{document}